\documentclass[sigconf]{acmart}
\usepackage{textcomp}
\usepackage{listings}
\usepackage[ruled,vlined]{algorithm2e}
\usepackage{algorithmic}
\newcommand{\upquote}{\text{\textquotesingle}}
\AtBeginDocument{%
  \providecommand\BibTeX{{%
    \normalfont B\kern-0.5em{\scshape i\kern-0.25em b}\kern-0.8em\TeX}}}

\setcopyright{acmcopyright}
\copyrightyear{2021}
\acmYear{2021}

\acmConference[DEBS '21]{DEBS '21: The 15th ACM International Conference on Distributed and Event-based Systems}{June 28--Juli 02, 2021}{Virtual Event, Milan, Italy}

\begin{document}


\title{The Synergy of Complex Event Processing and Tiny Machine Learning in Industrial IoT}

\author{Haoyu Ren}
\orcid{0000-0002-0241-6507}
\affiliation{%
  \institution{Siemens AG}
  \institution{Technical University of Munich}
  \city{Munich}
  \country{Germany}
}
\email{haoyu.ren@siemens.com}

\author{Darko Anicic}
\orcid{0000-0002-0583-4376}
\affiliation{%
  \institution{Siemens AG}
  \city{Munich}
  \country{Germany}}
\email{darko.anicic@siemens.com}

\author{Thomas Runkler}
\orcid{0000-0002-5465-198X}
\affiliation{%
  \institution{Siemens AG}
  \institution{Technical University of Munich}
  \city{Munich}
  \country{Germany}
}
\email{thomas.runkler@siemens.com}


\begin{abstract}
Focusing on comprehensive networking, big data, and artificial intelligence, the Industrial Internet-of-Things (IIoT) facilitates efficiency and robustness in factory operations. Various sensors and field devices play a central role, as they generate a vast amount of real-time data that can provide insights into manufacturing. The synergy of complex event processing (CEP) and machine learning (ML) has been developed actively in the last years in IIoT to identify patterns in heterogeneous data streams and fuse raw data into tangible facts. In a traditional compute-centric paradigm, the raw field data are continuously sent to the cloud and processed centrally. As IIoT devices become increasingly pervasive and ubiquitous, concerns are raised since transmitting such amount of data is energy-intensive, vulnerable to be intercepted, and subjected to high latency. The data-centric paradigm can essentially solve these problems by empowering IIoT to perform decentralized on-device ML and CEP, keeping data primarily on edge devices and minimizing communications. However, this is no mean feat because most IIoT edge devices are designed to be computationally constrained with low power consumption. This paper proposes a framework that exploits ML and CEP's synergy at the edge in distributed sensor networks. By leveraging tiny ML and micro CEP, we shift the computation from the cloud to the power-constrained IIoT devices and allow users to adapt the on-device ML model and the CEP reasoning logic flexibly on the fly without requiring to reupload the whole program. Lastly, we evaluate the proposed solution and show its effectiveness and feasibility using an industrial use case of machine safety monitoring.

\end{abstract}




\maketitle

\section{INTRODUCTION}
The Industrial Internet of Things (IIoT) refers to a system where hundreds of millions of industrial machinery, whether it is an engine on an airplane or a motor in a factory, are connected to the network. Sensors are installed on equipment to collect and share data over the network, monitoring and analyzing factory operation. The whole IIoT network generates a large amount of data in real-time, denoted as big data \cite{Mohamed2014}. The importance of big data depends not only on how much data are available but also on how the data are handled. The information needs to be analyzed and combined with background knowledge in a continuous, real-time manner to make quick decisions.

Complex event processing (CEP) is a widely deployed technique to process events and detect patterns from multiple heterogeneous streaming sources. The user can define logic rules to describe the desired sequence and pattern of primitive events. The CEP system then runs reasoning on low-level arriving events against the rules and produces complex (derived) events upon matching. However, with the growing size and complexity of IIoT networks, sequence matching in raw stream data becomes inefficient. It is bound to high cost since every CEP rule must be defined and examined by human experts. Moreover, not every pattern is rule-describable. For example, it is almost impossible to define rules for diagnosing machines' wear from pictures in visual inspection, although it is easy for a trained worker. 

Furthermore, many IoT applications require a higher level of awareness and reasoning over the current situation instead of simple combinations of primitive events. Machine learning (ML), especially deep learning, is one of the most innovative paradigms that can alleviate CEP's problem by learning data representations and identifying hidden information from transient raw events. While neural network (NN) models have been proved to deliver excellent inference results on instantaneous temporal data, they cannot infer the spatial and temporal contexts between atomic events that evolve over a long time, denoted as complex events. For example, it is hardly possible to use NN to describe a scenario that a machine's abnormal behavior will cause another machine's breakdown within one minute. 

Therefore, there is a need to combine CEP and ML to fully open up their potential. Currently, most ML and CEP solutions are specialized for platforms with considerable computing resources, e.g., powerful computing units, server and cloud-based services \cite{ASW21}\cite{apache21}\cite{tensorflow2015-whitepaper}\cite{NEURIPS2019_bdbca288}, where the data are first sent to the data warehouse by edge devices and then processed centralized,  which is referred to as the compute-centric paradigm. With such infrastructure, a tremendous amount of data from heterogeneously distributed IoT devices needs to be transmitted continuously over the network, which can saturate the bandwidth quickly and provide hackers with vulnerable attack vectors. 

As sensors and microcontrollers (MCUs) become cheaper, more powerful, and pervasive, more and more manufacturing professionals see edge computing as the next opportunity for the IIoT. With edge computing, the data processing is shifted and distributed to the sources where the data are generated, directly on edge IIoT devices, which is denoted as the data-centric paradigm. By edge IIoT devices, we mean MCUs with limited resources, e.g., less than 0.1W power consumption, 64MHz CPU frequency, and 256KB RAM. By enabling ML and CEP's synergy in a distributed manner, directly at sources, we can keep data mainly on edge devices, which saves communication costs, improve security, reduce latency, and preserve data privacy without letting sensitive raw data leave devices. It offers the benefit of real-time distributed data analysis for faster decision-making and more intelligence in the complex production environment, thereby creating countless opportunities for industrial applications. However, edge IIoT devices are designed to live long with limited resources and low power consumption. Most of them are bare metal and do not even have an operating system. This poses difficulties for applying ML and CEP on them to realize decentralized sensor networks. 

To tackle the challenges mentioned above, we designed a framework that combines NN with CEP to perform inference and reasoning on distributed IoT devices with limited computing resources:

\begin{itemize}
    \item Several light-weight inference libraries \cite{Tianqi2018}\cite{Robert2020}\cite{STM21} have already been developed for NN inference on bare-metal MCUs. However, none of them supports on-device post-training, which impedes the flexibility of IIoT as the actual deployment environment may change constantly, and the performance of the pre-trained model may decrease. Thus, we built a TinyOL system \cite{Ren2021} based on Tensorflow Lite Micro \cite{Robert2020} which can run NN inference on MCUs and improve the pre-trained model using field streaming data under an incremental online learning setting. 
    
    \item We introduced the micro CEP engine derived from the reasoning system {\em ETALIS} \cite{DarkoAnicic2012}, which can continuously match incoming events against user-defined patterns directly at the edge with low latency and high throughput. The instantaneous inference results made by the TinyOL, together with other atomic events, can be fed into the on-device micro CEP engine for searching a candidate complex event. Vice versa, the output of the engine can be used as pre-processed data for the NN model. Once the engine recognizes events in the matching sequence, the results are emitted immediately.  Moreover, the reasoning topology can be easily changed by injecting CEP rules into the system at runtime.
\end{itemize}

The framework is evaluated in an industrial scenario of machine safety monitoring. We demonstrate its flexibility and versatility by incorporating convolutional NN, autoencoder NN \cite{Ren2021}, and CEP on two Arduino Nano 33 BLE boards \cite{Arduino21}. Besides, the system's statistical performance is reported concerning memory consumption, processing speed, and throughput. We show that the framework is light-weighted to be fit into constrained MCUs. More importantly, it is capable of handling streaming data with high inference speed, which proves its usability in most IIoT applications.

The remaining work is structured as follows. The next section, Section \ref{section:requirement}, discusses the requirement of system development from the industrial perspective. Section \ref{section:related_work} summarizes the literature reviews regarding CEP, TinyML \cite{tinyML2021}, and some industrial applications. This is followed by Section \ref{section:approach}, which includes the system design of our proposed solution. In Section \ref{section:evaluation}, the framework is evaluated using an industrial safety monitoring use case, and its performance on an Arduino board is reported. Lastly, Section \ref{section:conclusion} provides the conclusion and potential areas for future works.

\section{REQUIREMENT ANALYSIS}
\label{section:requirement}

This section describes the motivation for building the proposed system and the requirements that need to be satisfied against the industrial background.  

\subsection{Requirement}

\subsubsection*{The Local Device Intelligence}

Data, especially shop floor data, produced and consumed by industrial sensors and actuators are local by its nature. It is a requirement to compute data closer to the location where it is produced (e.g., on-device) to improve response times and save bandwidth. So far, CEP and ML have been applied on non-constrained computers or in the cloud. We are now pushing the intelligence provided by these technologies to the field. However, industrial field devices are constrained. It is a challenge to enable a synergy of CEP and ML to work on constrained IoT devices.

\subsubsection*{The Distributed Computation}

As the network expands and the number of devices increases, the overall decision-making process in IIoT becomes a distributed computation problem. The decisions derived by one sensor node may be influenced by the data and the contextual information from other nodes. Thus, it is our goal to consider the synergy of CEP and ML in a decentralized environment. 

\subsubsection*{The Energy Consumption}

Most IoT devices are power-constrained objects, as they are designed to live long with minimal energy consumption. It is essential to consider their hardware limitation while constructing applications for the IIoT system.

\subsubsection*{The Flexibility}

Nevertheless, to keep pace with changing customer demands, it is important to provide IIoT networks with flexible configuration features.  In the current solution, static ML models \cite{Robert2020} or complex event processors with predefined rules \cite{Proctor2012}\cite{Lars2007} are flashed to the device before the runtime. To change the underlying program behaviors, e.g., configuring the reasoning logics or updating the on-device ML models, a heavy firmware update is unavoidable, which makes the deployment of IIoT in the industry environment a challenging task:

\begin{itemize}
    \item Different working conditions have different characteristics. A predefined factory setting does not work for all. Often, the fine-tuning of rules and models is a necessary step before deployment.
    \item Some customers want to update the program more frequently in response to the developing requirements. To perform the firmware update, the current production activity has to be suspended. Considering the enormous amount of IoT devices, updating the firmware in each of them is costly.  
\end{itemize}

\subsection{Motivation}

By shifting the cloud processing to the field devices, we can enable edge computing and turn centralized IIoT into decentralized sensor networks. The data are processed at the source locally, which reduces latency. More importantly, data analysis can be performed directly in the field without relying on an external connection, which guarantees data privacy because storing data in a distributed manner is safer than storing it in a single location (such as the cloud). Only necessary information is sent to the cloud, saving energy and bandwidth. Combined with the widespread and low-cost of edge IIoT devices, this opens up countless industrial applications opportunities because they can provide customized and flexible solutions in the complex production environment.

This work presents a light-weight framework combining on-device ML and CEP reasoning targeted at constrained IIoT devices with maximized flexibility and minimized power consumption in mind. Using the TinyOL system in the framework, the user can adapt the pre-trained NN on the device by learning from the field data and running inference afterward. Another building block of the framework, the micro CEP engine, can analyze streaming events among sensor networks, including the NN's inference results, and match them against the rules in real-time. The reasoning patterns can be changed at any time by pushing rules into the engine without rewriting the device's whole firmware. The system enables the user with little programming experience to flexibly manage and control an IIoT system.

\section{RELATED WORK}
\label{section:related_work}

Complex event processing is a well-established field applied across various domains. A CEP system addresses sensor networks' ubiquity by efficiently matching input streaming events against a pattern, where irrelevant incoming data can be discarded immediately. The survey \cite{Lajos2010} discussed the early works of research, implementation, terminology, applications, and open issues in CEP.  Several commercial and open-source event processing tools have been developed in recent years \cite{apache21}\cite{TIBCO21}\cite{ASW21}.  The work \cite{MarcosDiasdeAssuncao2018} studied state of the art on CEP mechanisms and presented their drawbacks in heterogeneous IoT environment. The survey \cite{Giatrakos2020} discussed the techniques, the opportunities, and the challenges of CEP in the big data era.

 While most CEP tools are designed for centralized cloud analysis, works like \cite{Pablo2018} try to extend the CEP to mobile devices to leverage decentralized architectures.  In \cite{NithyashriGovindarajan2014}, a by-partitioning CEP pipeline is proposed to use both edge and cloud resources for stream processing. A micro-service-based method is introduced in \cite{MiraVrbaski2018} to handle CEP in IoT systems. A few works \cite{Lina2019}\cite{ChingyuChen2014} focus on a hierarchical complex event model to adapt CEP engine in distributed sensing environment. Although these solutions proved to work in IoT systems, they do not work on the sensor networks composed of constrained MCUs. In their works, the IoT devices are usually Raspberry Pi-class with far more available resources. This class of devices does not fit into the scope of constrained IoT devices. Besides this, they assume the logical predicates do not need to be updated after deployment and are thus flashed to the devices prior to the runtime as static rules, which restrict the CEP engine's versatility in the context of IIoT applications.

With the rise of AI, engineers are paying more and more attention to applying ML in the industrial field, especially in IIoT. As an emerging AI sub-field, TinyML is dedicated to providing neural network-based solutions at the edge. Impressive results have been achieved recently, such as tinyML algorithm \cite{MohammadRastegari2016}\cite{SonHan2015}, hardware \cite{Jinook2019}\cite{Bert2018}, and application \cite{SanchezIborra2020}\cite{Miguel2020}. Several libraries are developed to support NN inference at the edge, including Google's TFLite Micro \cite{Robert2020}, Arm's CMSIS-NN \cite{LiangzhenLai2018}, Apache's TVM \cite{Tianqi2018}, and STM's X-CUBE-STM AI \cite{STM21}. However, these libraries assume that the NN is first trained on a powerful machine then flashed to the MCU without supporting on-device training and fine-tuning, which might lead to the deterioration of NN's performance in the field after deployment\cite{Ren2021}.

Combining ML with CEP is a promising solution to circumvent some technical limitations in the IIoT. In work \cite{JonasWanner2019}, the author surveyed the synthesis of both paradigms and their transferability to intelligent factory use cases. Some researchers tried to derive CEP reasoning rules using ML methods \cite{Nijat2015}\cite{RaefMousheimish2017}\cite{RalfBruns2019}.  A Bayesian network is used to predict incoming online events in work \cite{YonghengWang2018}. Many CEP and ML-based solutions have been applied in the real world. A CEP and ML-based approach to support fault-tolerance of IoT systems is proposed in work \cite{Power2019}.  In \cite{Tianwei2019}, a framework based on CEP and deep learning is implemented, and an unattended bag computer vision task is illustrated to evaluate its feasibility. Unlike other works, CEP is used to schedule distributed ML training on Raspberry Pi in \cite{JoseAngelCarvajalSoto2016}. Nevertheless, many implementation designs depend on the usage of the cloud for communication.  None of those touch the area of constrained devices and test their approaches under industry settings.  

To sum up, although many advances have been achieved in CEP and ML, there exists a gap in constrained devices. Several questions remain unanswered: How to realize CEP and ML on tiny IIoT devices? How to maximize their potential and flexibility?  How to process the ubiquitous events stream against the industrial background? We now introduce our framework that combines the TinyOL system and the micro CEP engine to achieve flexible, decentralized IIoT on constrained field devices. 

\begin{figure}[htbp]
  \centering
  \includegraphics[width=\linewidth]{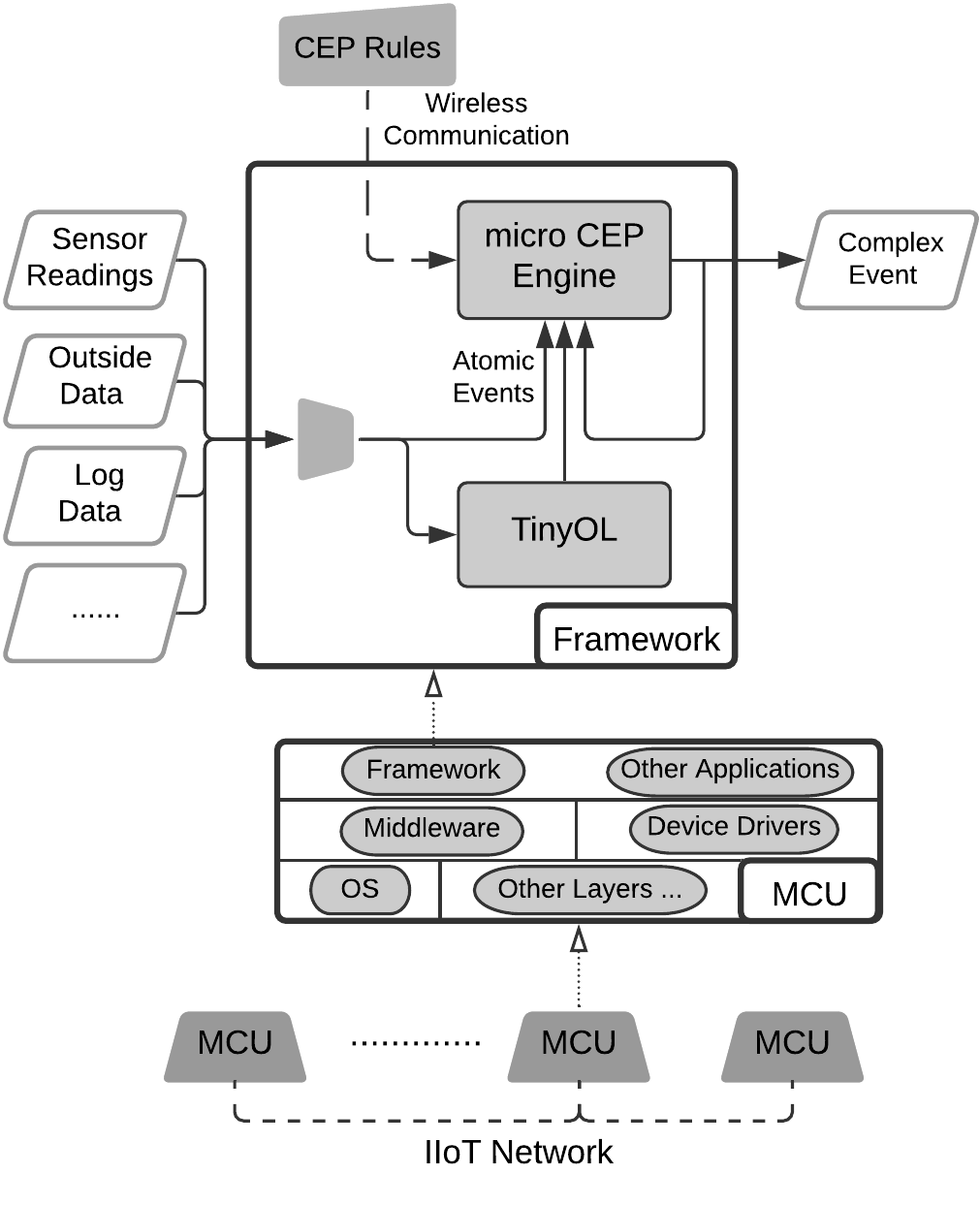}
  \caption{System Overview of the Framework}
  \label{figure_1}
\end{figure}

\section{APPROACH}
\label{section:approach}

Our proposed framework consists of two building blocks: the TinyOL system for powering NN inference and the micro CEP engine for event processing and stream reasoning.  Such a framework can be installed as an embedded application in every tiny device inside the IIoT network to achieve NN and CEP's synergy at the field level. The system design is depicted in Figure \ref{figure_1}. 

Inside the TinyOL system, depending on the available resources of MCUs and onboard sensors, a certain number of prebuilt NN models are at the user's disposal. For instance, the framework can support up to three NN models on a MCU with 256 KB RAM and 1MB Flash.  The raw data are fed into the associated NN, and the corresponding inference results are instantaneously produced toward the specific task. 

Simultaneously,  measurements from other sources (log data, outside information, and sensors) are generated.  These data, including the NN's inference results, are identified as raw measurements that can be formatted and enriched into simple atomic events using their related information (e.g., the context, the timestamp of emitting, and the timespan that the measurement takes place, and various numerical values). The formed events then enter the micro CEP engine in the order in which they appear. The micro CEP engine automatically checks the temporal and spatial constellations between the events upon their arrival. Once the engine sees incoming events for a matching pattern, a result (complex event) is generated immediately. The emitted complex event can again be sent into the micro CEP engine as a "simple event" of a new type for composing more complex events against user-defined rules. During the runtime, the logical rules can be updated and injected into the engine via communication protocols, e.g., Bluetooth Low Energy (BLE), WiFi.

The whole system performs in real-time as the measurements take place, which effectively improves the flexibility and analysis capability. In the following, we introduce the system design of the TinyOL system and the micro CEP engine.

\subsection{The TinyOL System}

\begin{figure}[t]
  \centering
  \includegraphics[width=\linewidth]{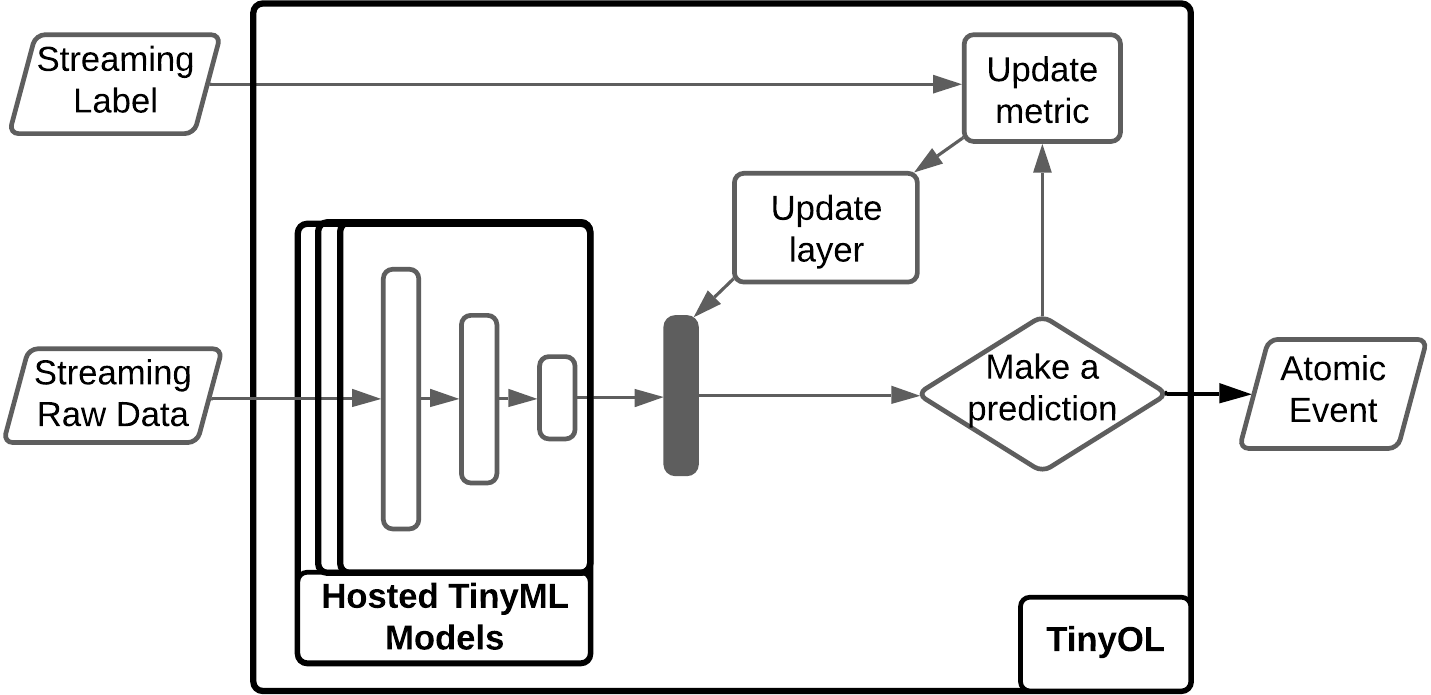}
  \caption{Overview of the TinyOL System}
  \label{figure_2}
\end{figure}

The TinyOL system supports neural network inference on constrained MCUs with Tensorflow Lite for Microcontroller \cite{Robert2020} under the hood. The system design is shown in Figure \ref{figure_2}. Here, only the conceptual idea behind the building block is introduced. For greater detail, please refer to work \cite{Ren2021}. 

\begin{algorithm}[t]
  \caption{Workflow of the TinyOL System}
  \begin{algorithmic}[1]
    \FOR{x in StreamingData} 
        \STATE Choose the corresponding TinyML model
        \STATE x\upquote \ = TinyML.PreProcess(x);
        \STATE y\upquote \ = TinyOL.Inference(x\upquote);
        \IF{label y is available}
          \STATE TinyOL.UpdateMetrics(y\upquote, y); 
          \STATE TinyOL.UpdateWeights(y\upquote, y);
        \ENDIF
    \ENDFOR
  \end{algorithmic}
  \label{a1}
\end{algorithm}

Current market solutions for powering NN on MCU do not support on-device training \cite{TFLiteMicro21} due to the lack of available resources. This results in difficulties in the industrial environments, e.g., model performance degeneration, rigid deployment, and high maintenance cost. TinyOL system enables the on-chip post-training by leveraging the idea of incremental online learning.  While data are streaming in, the TinyOL can not only run inference but also update the corresponding NN model by learning from the data one at a time. To limit the computational complexity, the first part of the pre-trained NN is frozen and stored in flash as a C array, and the fine-tuning of the last few layers takes place in RAM, which works similarly to transfer learning. The trainable layers are marked in grey in Figure \ref{figure_2}, which can be initialized, updated, and customized. The pseudo algorithm of TinyOL is illustrated in Algorithm \ref{a1}. For each incoming sample, the system first runs a prediction against a NN model from the hosting model pool. Subsequently, the trainable layers' weights in the model get updated using online gradient descent if a label is provided.  As the prediction and training are interleaved, the data can be discarded after each round. Thus only one sample remains in the system, whereas all the historical data need to be saved in the traditional offline/batch training architecture. 

The TinyOL system can help the user to adapt the prebuilt NN models to their specific working environment. The prediction outputs from TinyOL are wrapped with other semantic information (e.g., name, inference time, scores) into atomic events and sent to the micro CEP engine for further processing.

\begin{figure}[t]
  \centering
  \includegraphics[width=\linewidth]{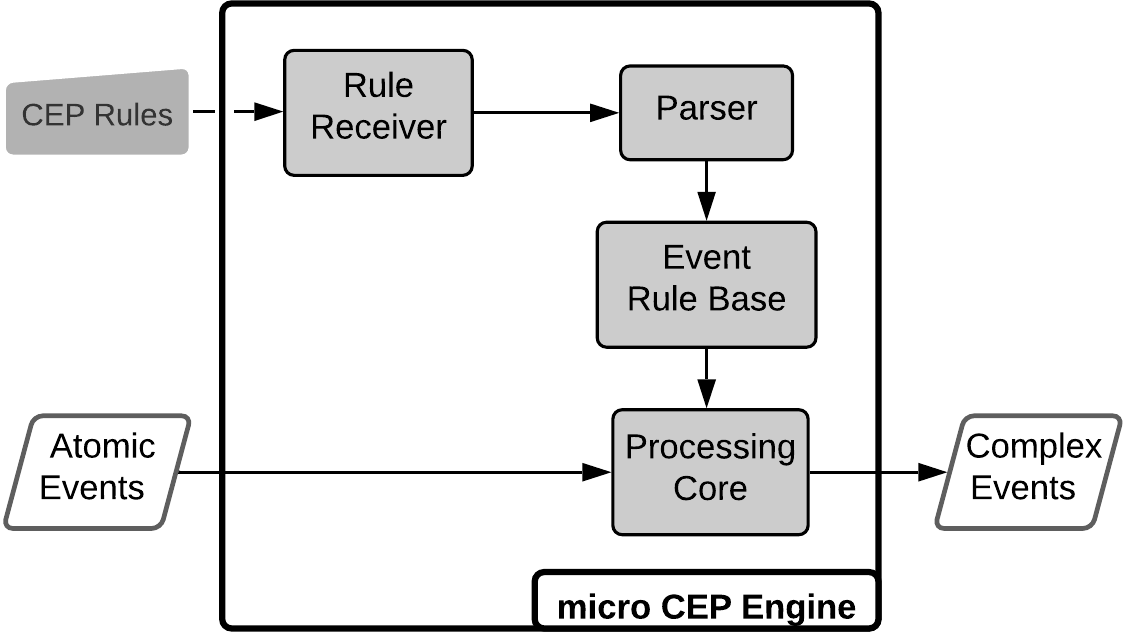}
  \caption{System Overview of the micro CEP Engine}
  \label{figure_3}
\end{figure}

\subsection{The Micro CEP Engine}

The micro CEP engine is a light-weight system designed for power-constrained IIoT devices to deal with sensor networks' ubiquity. As the IIoT network gets more complex and distributed, the ability to quickly respond to increasing streaming data and changing trends becomes a decisive factor in a company's success. The system design of the micro CEP engine is shown in Figure \ref{figure_3}. The event engine consists of three parts, a rule receiver accepting and managing the CEP rules provided by the user on the fly, a parser used to interpret the rules, a processing core to process incoming events against the rules and return the results.

The language that powers the engine is derived from {\em ETALIS} \cite{DarkoAnicic2012}.  Before the streaming raw data are fed into the engine, it is converted into atomic events. An atomic event is the most simple data representation in the language where the raw data is assigned with a set of attributes that refine its semantic meaning, such as the name, the time, the measurement value, and the unit.  An atomic event is defined as follows:

\begin{verbatim}
  event_name[start_time, end_time](arguments),
\end{verbatim}

where the $event\_name$ is a string to identify the event, the $start\_time$ and the $end\_time$ indicate the timestamp of the start and the end of the event in milliseconds, and the $arguments$ contain the comma-separated information of the event. As an example, a measurement taken from a temperature sensor can be written as:

\begin{verbatim}
  temperature_event[2000, 2200](24, Celsius),
\end{verbatim}

where $2000$ and $2200$ indicate that the measurement starts at timestamp $2000\ ms$ and ends at timestamp $2200\ ms$, and $24$ is the temperature value. $Celsius$ is the unit of the measurement, which is optional information for enriching the event with semantic information, e.g., from a Knowledge Graph \cite{hogan2021}.  

Given the notation of event, we can formalize the definition of the rules for the micro CEP engine. The most straightforward rule can be defined as follows:

\begin{verbatim}
  complex_event[start_time, end_time](arguments) :- 
  atomic_event[start_time, end_time](arguments),
\end{verbatim}

where a $complex\_event$ is generated whenever an $atomic\_event$ is pushed into the engine, the timestamp and measurement information can be inherited from atomic to the complex event. 

Filter conditions can be integrated into the rules by using $where$ clauses. For instance, the following rule can filter out the temperature events with a value below $20$ degrees Celsius:

\begin{verbatim}
  filtered_temperature[_,_](X) :- 
  temperature_event[_,_](X, Celsius) where(X>20),
\end{verbatim}

where the variable $X$ is used to capture numerical values from the atomic to the complex event, and $\_$ is a placeholder (unnamed variables as in Prolog-like languages) for timestamps. 

Once a rule is triggered, a complex event is formed and ejected by the engine, which summarizes the correlations between triggering events and the underlying rules. For example, if we use the filtering rule mentioned above and push the previous temperature event into the engine, the following complex event will be triggered:

\begin{verbatim}
  filtered_temperature[2000, 2200](24, Celsius).
\end{verbatim}

We can see that the complex event has the same form as the atomic event, which can be again pushed into the engine to form more complex events.

Several operators are supported to compose and express more complicated event patterns:

\begin{itemize}
\item $and$ : Conjunction operator.
\item $seq$ : Sequence operator.
\item $or$ : Disjunction operator.
\item $nseq$ : Negation operator.
\item $kseq$ : Kleene closure operator.
\item $lambda$ : Aggregation operator.
\end{itemize}

We focus on $and$ and $lambda$ operators in this paper since other operators share similar patterns. For details about the other event operators from the $ETALIS$ language, an interested reader is referred to \cite{DarkoAnicic2012}. The conjunction operator $and$ is used to detect the occurrence of events regardless of the order. Additionally, we can impose temporal constraints by assigning a specifier $count$ or $range$ in the square bracket after the rules, which states the type and the length of the valid sliding window. For example, we can use the following rule to match the pattern that event $A$ and event $B$ happen within $5$ seconds, no matter in which order they appear. Additionally, the $where$ condition constrains an occurrence of event $A$'s to have a value smaller than the one from an occurrence of the event $B$:

\begin{verbatim}
  complex_sequence[_,_](X, Y) :- eventA[_,_](X)
  and eventB[_,_](Y) where(X<Y) [range 5 s].
\end{verbatim}

The aggregation operator $lambda$ can be used to apply aggregation functions over event streams. These functions are currently supported in the aggregation operator:

\begin{itemize}
\item $abs()$ : Absolute value
\item $sum()$ : Summation function
\item $avg()$ : Average function
\item $min()$ : Minimal function
\item $max()$ : Maximal function
\end{itemize}

For example, the following rule is used to produce the average temperature over last five measurements:

\begin{verbatim}
  temp_avg[_,_](Y) :- lambda { temperature_event
  (X, Celsius), *, Y := avg(X) } [count 5].
\end{verbatim}

The CEP engine provides a running buffer to store relevant events against the rule. Once an event turns out of the scope, it will be discarded efficiently. Besides the functionalities introduced above, another key feature of the micro CEP engine is the configurable reasoning rules at runtime. The user can update the rules towards the desired information and reasoning topology at any time by injecting them via wireless communication protocol like WiFi and BLE.

\section{EVALUATION}
\label{section:evaluation}

This section first explains our approach's practical usability throughout a case study of safety monitoring in a factory. Later on, we validate the system and the use case with a simplified setting using two Arduino boards \cite{Arduino21}, one camera sensor \cite{camera2021}, and a ventilator. Finally, we report the framework's performance, including processing time, memory consumption, and throughput.

\subsection{Use Case - Concept}

\begin{figure}[htbp]
  \centering
  \includegraphics[width=0.95\linewidth]{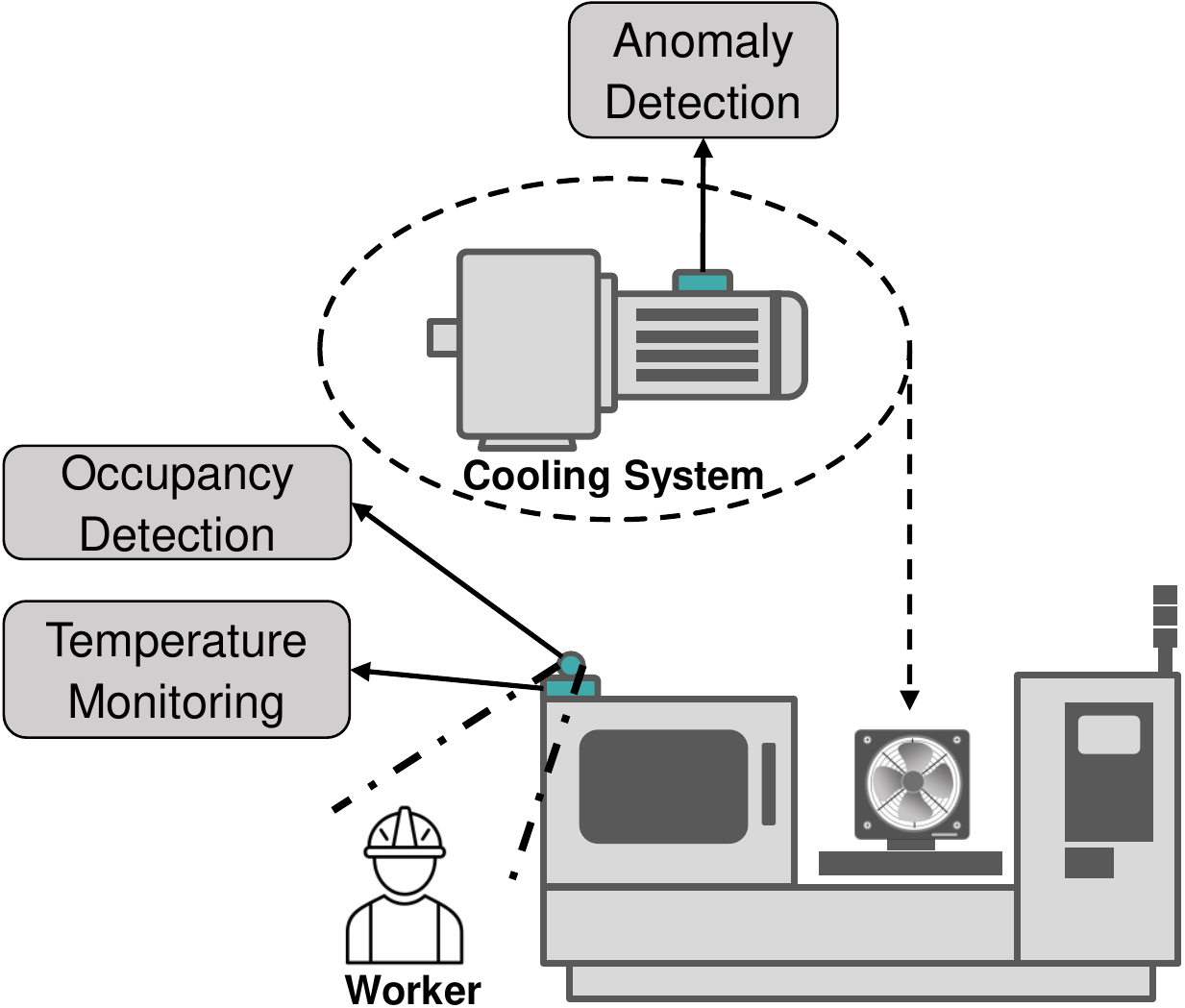}
  \caption{Scenario: Safety Monitoring in Factory}
  \label{figure_4}
\end{figure}

Assume a rotating machine running in a factory. To keep it operating under the best condition, in this case, we refer to under optimal environment temperature, we monitor the cooling system's behavior, the temperature in the plant, and the worker's occupancy, as depicted in \ref{figure_4}. These activities compose of a series of atomic events, which can be easily ingested using our framework. For this case study, we use two tiny IoT devices with the framework embedded, as marked 
as the small rectangles in the figure. 

The first IoT device is attached to the cooling system's motor to detect its abnormality in real-time. We choose a general pre-trained autoencoder NN from the NN pool hosted on the device to detect the machine's anomaly in the vibration data collected by the onboard accelerometer.  Before the deployment, we first fine-tune the autoencoder towards the new working environment using the TinyOL system. By learning from the field data, we adapt the general anomaly detection model to fit the specific machine and enhance its prediction performance. During the regular operation, the autoencoder generates an anomaly score event at each inference to indicate if the cooling system is working normally. Whenever the anomaly score exceeds the threshold, a warning event is generated and forwarded to the second IoT device.

The second IoT device is responsible for checking if a worker is on-site using computer vision when a warning event from the first device is received. Each time, a picture is captured by the camera and processed by the on-device convolutional NN (CNN) model, and an occupancy event based on the inference results is generated. If a worker is detected in the camera field, the second IoT device lights up the warning LED without further action. If no worker is currently near the machine, the IoT device establishes a connection with the cloud and sends an alarm event to the backend. 

We can also update the monitoring strategy by altering the rules in the micro CEP engine at runtime. Assume the indoor temperature increases when the cooling system breaks down, which dangers the machine's operation. We want to improve the intelligence of the system by integrating temperature monitoring. A rule can be added to start a backup cooling system if (1) the cooling fan operates in an anomalistic manner,  (2) no on-site worker event is presenting, and (3) the temperature event exceeds the threshold,

Now, we can use our framework to elaborate on how the scenario above can be solved efficiently. For the first IoT device, we define an incoming event to be monitored as follows:

\begin{verbatim}
  1.1 anomaly_score[t1, t2](x),
\end{verbatim}

where every anomaly event contains the timestamp of the inference as well as the anomaly score from the NN, and the number $1$ before the rules and events means that this action takes place in the $first$ device.

Now, an average aggregation rule is created to calculate and smooth the anomaly score over the measurements from the last 10 seconds for filtering out some outliers, e.g., vibration intervention from outside.

\begin{verbatim}
  1.2 smoothed_anoamaly_score[_,_](Y) :- lambda 
       { anomaly_score(X), *, Y := avg(X) } [range 10 s].
\end{verbatim}

We set another rule with a threshold of 1 for creating warning events:

\begin{verbatim}
  1.3 warning[_,_](X) :- 
      smoothed_anomaly_score[_,_](X) where(X>1).
\end{verbatim}

If a warning event is presented, it will be forwarded to the second IoT device for activating person detection:

\begin{verbatim}
  1.4 if warning[_,_](Y) do
        sending a request for occupancy detection.
\end{verbatim}

Since running a computer vision task is power consuming, we only activate person detection if it is confident that something is wrong in the cooling system. In this case, that is the second device receives a warning event. This mechanism is also known as a cascade system. We again define the following atomic event in the second device to represent the results of CNN's occupancy detection. 

\begin{verbatim}
  2.1 if warning[_,_](Y) do
        occupancy_score[t1, t2](x),
\end{verbatim}

The number $2$ before the rules and events means that this action occurs in the $second$ device. The decision boundary of the CNN's output is chosen to be 0, where a positive value means a worker is detected in the camera field and vice versa. Based on this, We construct two complex event patterns to indicate whether a person is near the machine:



\begin{verbatim}
  2.2 not_occupied[_,_](X) :- 
      occupancy_score[_,_](X) where(X<0),
\end{verbatim}

\begin{verbatim}
  2.3 occupied[_,_](X) :- 
      occupancy_score[_,_](X) where(X>0).
\end{verbatim}

If a $not\_occupied$ event shows up, an alarm message should be sent to the monitoring center via the cloud. Otherwise, only the onboard LED will be light up to warn the on-site worker.

\begin{verbatim}
  2.4 if occupied[_,_](X) do
        lighting up the warning lamp,
      else if not_occupied[_,_](X) do: 
        sending an alarm to the cloud.
\end{verbatim}

After the experts in the monitoring center received the alarm, they want to proactively integrate the temperature monitoring and a backup cooling system into the IIoT system in case there is no on-site worker available. The following reasoning rule can be added to the second device on the fly, where the backup cooling system is turned on if the environment temperature becomes too high (higher than 30 °C) and no worker is presented:

\begin{verbatim}
  2.5 backup[_,_](Y) :- temperature[_,_](Y) and 
      not_occupied[_,_](X) where (Y>30) [range 1 s].
\end{verbatim}

The $range$ clause is used to constrain the occurring interval between the two events. 

\begin{figure*}[htbp]
  \centering
  \includegraphics[width=0.75\textwidth]{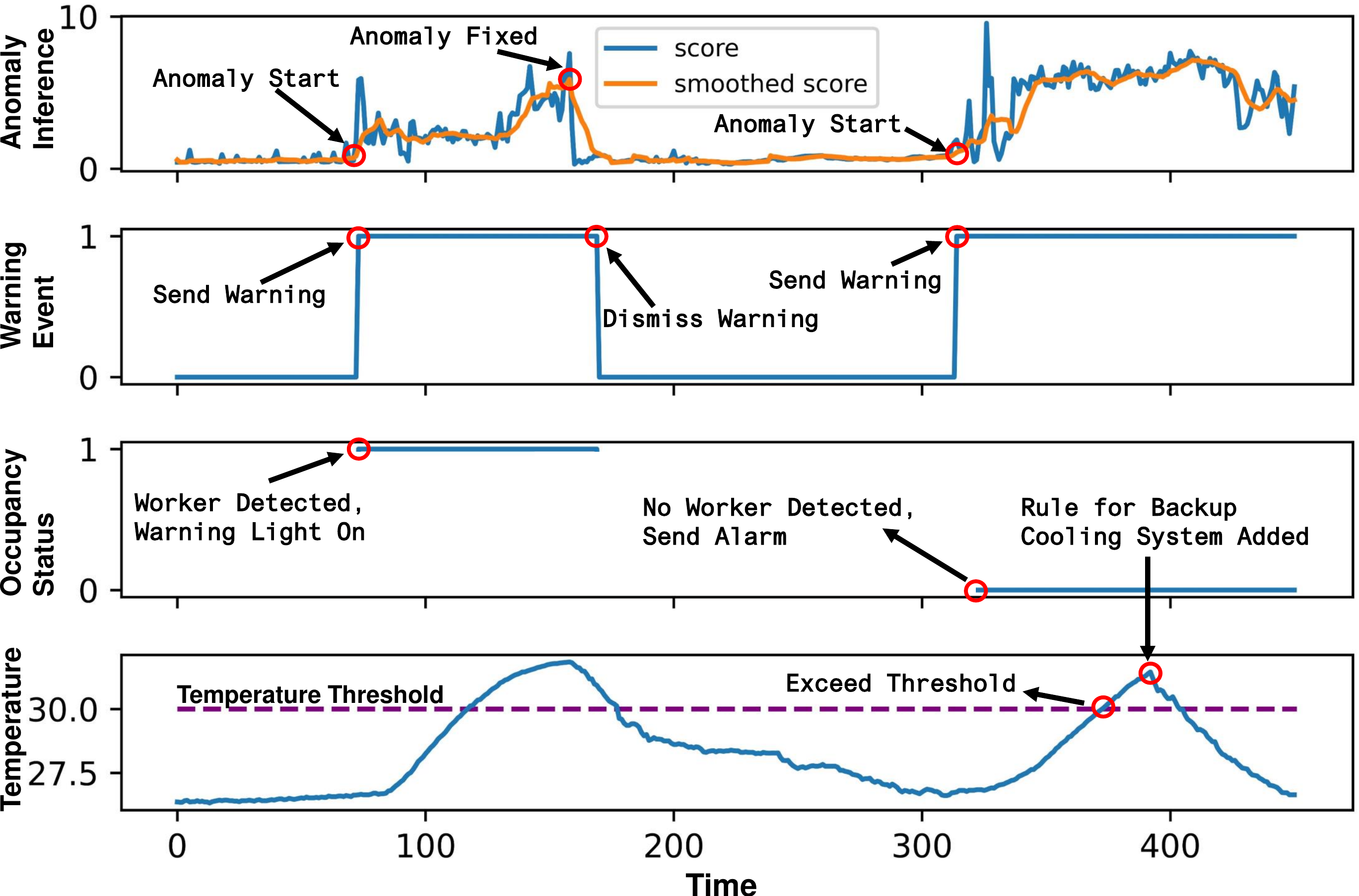}
  \caption{Process Monitoring of the Use Case.}
  \label{Experiment}
\end{figure*}

\subsection{Use Case - Experiment}

Next, we examine the use case using two Arduino Nano 33 BLE boards \cite{Arduino21}, an OV7675 image sensor \cite{camera2021}, and a cooling fan under a simplified setting. 

\begin{figure}[htbp]
  \centering
  \includegraphics[width=0.9\linewidth]{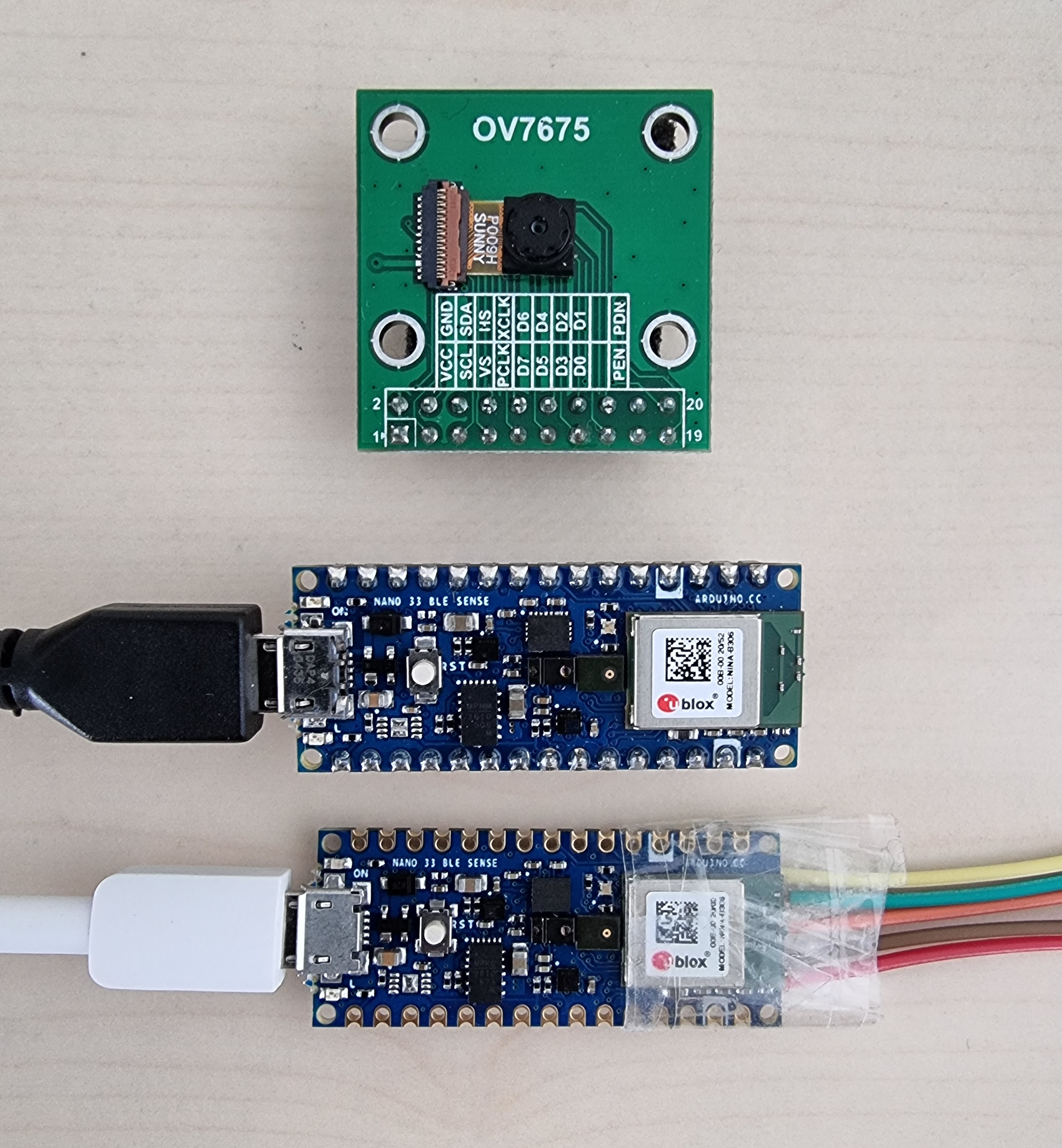}
  \caption{The Arduino boards and the OV7675 Camera.}
  \label{hardware}
\end{figure}

The board is featured with 256KB SRAM, 1MB flash, and powered by a Cortex™-M4 CPU running at 64 MHz, which has a similar computational capability as the industrial IoT devices (industry-level of reliability and durability is not considered in this paper). The low-voltage image sensor OV7675 supports up to VAG image resolution (640*480), which can operate at up to 30 fps in a small footprint package. The hardware used in the experiment is depicted in Figure \ref{hardware}.

The first board is placed on the cooling fan, which emulates monitoring the cooling system.  The second board is equipped with a camera for person detection. 

Before we put the system online, we first fine-tune the autoencoder in the first board using the TinyOL system. No labels need to be provided as it is an unsupervised anomaly detection NN.  For more implementation detail of the autoencoder, please refer to \cite{Ren2021}. 

Subsequently, we activate the framework on both boards to start the experiment. Upon starting, several parameters are under monitoring over time, as shown in Figure \ref{Experiment}. 

The first subplot in the figure illustrates the inference score of the anomaly detection and its smoothed outcome from the first device, which are the value of the event 1.1 and the result of rule 1.2. With rules 1.3 and $1.4$, the first IoT device sends a warning to notify the second IoT device when the smoothed score exceeds the pre-defined threshold 1.0. As shown in the second subplot, the value 1 of the warning event means the warning is sending, whereas the value 0 implies the cooling system works normally. One can observe in the plot that the cooling system experienced anomaly activity twice in the experiment. However, only the first anomaly gets fixed. 

With rule 2.1, the person detection CNN on the second device will be activated to check workers' presence once a warning is received. By combining rules 2.2, 2.3, and 2.4, the person occupancy status is monitored. As depicted in the third subplot, the value 1 refers to a worker is detected. Hence, the warning light is turned up. Otherwise, an alarm should be sent to the monitoring center. That is, no worker is on-site with the occupancy status of value 0.  We can see in the plot that there are on-site workers during the first incident, but no person is detected at the second anomaly. That is the reason why the second incident did not get fixed quickly. 

In the fourth subplot, we draw the changing temperature in the plant influenced by the cooling system's working condition. During the cooling system's anomaly, the environment temperature rises, which puts the factory operation at risk. Therefore, the engineers in the monitoring center add a new rule 2.5 remotely to the second IoT device for turning on the backup cooling system when the temperature is higher than the threshold of 30 °C, and no worker is nearby. After the rule injection, one can see that the backup cooling system is started and the temperature decreases. 

The use case illustrated above demonstrates the versatility and the flexibility of the framework in a decentralized network with a simplified setup. Depending on the type of sensors, actuators, and the concrete use cases, we can easily customize the reasoning topology of the whole IIoT, thanks to the framework. 

\subsection{Performance Evaluation}

Our framework is implemented in C/C++. We show its performance regarding memory usage, inference time, and processing ability in this section. The test is conducted on an Arduino Nano 33 BLE board. 

\begin{table}[b]
\caption{Core Runtime of the Systems.}
\label{t1}
\begin{tabular}{@{}lcc@{}}
\toprule
System           & \multicolumn{1}{l}{RAM {[}byte{]}} & \multicolumn{1}{l}{Flash {[}byte{]}} \\ \midrule
TinyOL System    & \char`\~7K                             & \char`\~135K                                \\
micro CEP Engine & \char`\~17K                              & \char`\~34K                                 \\ \bottomrule
\end{tabular}
\end{table}

Table \ref{t1} shows the memory usage of the core runtime of the systems. Both parts of the system (CEP and ML inference) are designed to run on power-constrained microcontrollers. As the results show, their runtime can easily fit in devices with limited resources.

\begin{table}[b]
\caption{Average Memory Consumption and Inference Speed of the NN Models Used in the Case Study.}
\label{t2}
\resizebox{\linewidth}{!}{%
\begin{tabular}{@{}lccc@{}}
\toprule
TinyML Model            & \multicolumn{1}{l}{RAM {[}byte{]}} & \multicolumn{1}{l}{Flash {[}byte{]}} & \multicolumn{1}{l}{Inference Speed {[}ms{]}} \\ \midrule
Anomaly Detection & 21120                              & 53616                                & 1.7                                          \\
Person Detection  & 45776                              & 271664                               & 1242.0                                         \\ \bottomrule
\end{tabular}%
}
\end{table}

Table \ref{t2} compares how much resources different TinyML models need and how fast the inferences rum under the TinyOL system. The results do not count the hardware libraries, e.g., the library to support sensor reading.  One can notice that the CNN/person detection model requires far more computing resources than the autoencoder/anomaly detection model. Therefore, a cascade system may be helpful in the application to reduce energy consumption: a light-weight ML model runs most of the time, not very accurate but fast reacting to the sensor readings, and a powerful yet heavy model can be activated when we need more precise and confident results.

\begin{figure}[t]
  \centering
  \includegraphics[width=1\linewidth]{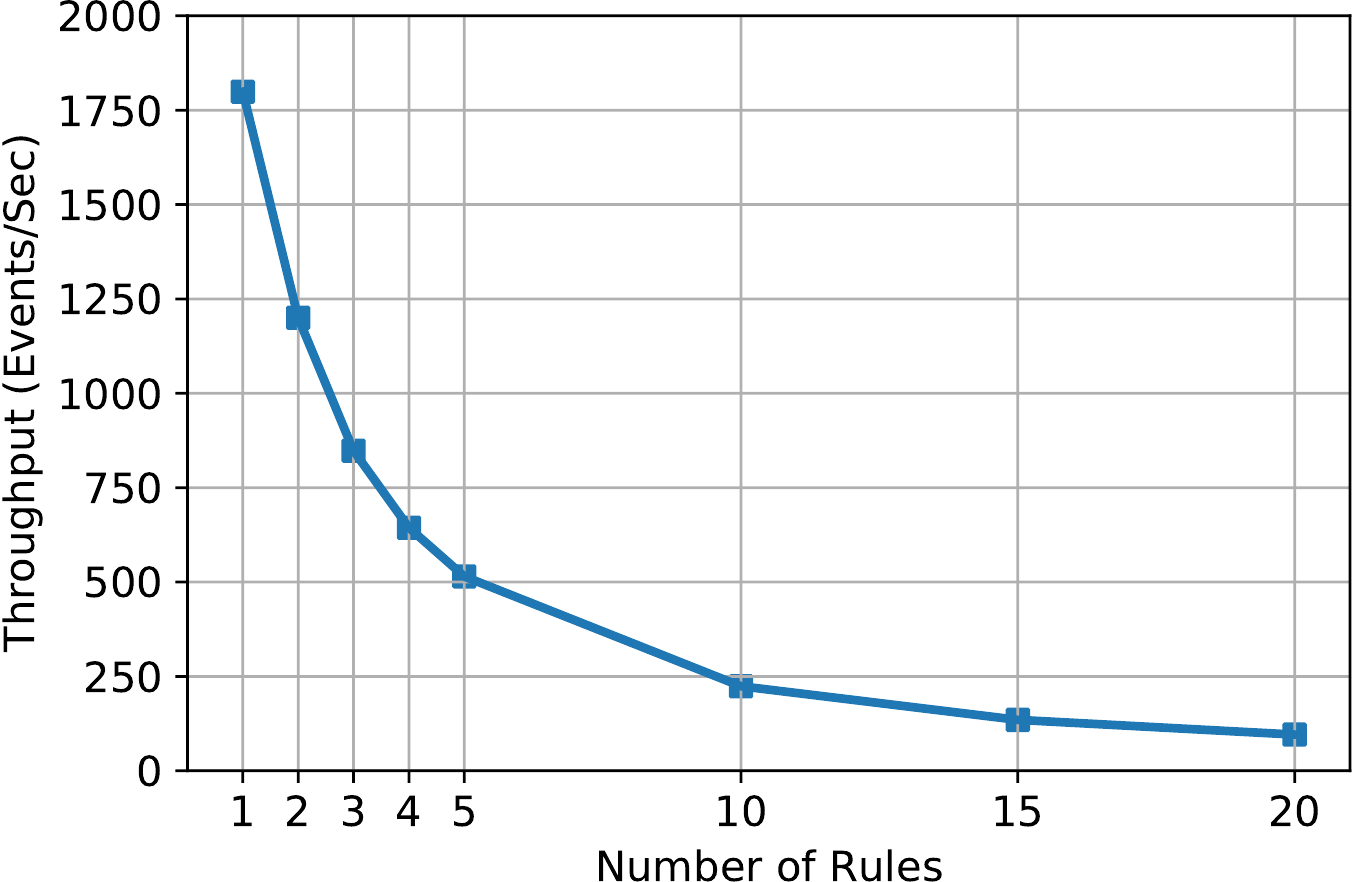}
  \caption{Throughput Test of the micro CEP Engine (Throughput vs Number of Rules).}
  \label{Throughput_1}
\end{figure}

\begin{figure}[t]
  \centering
  \includegraphics[width=1\linewidth]{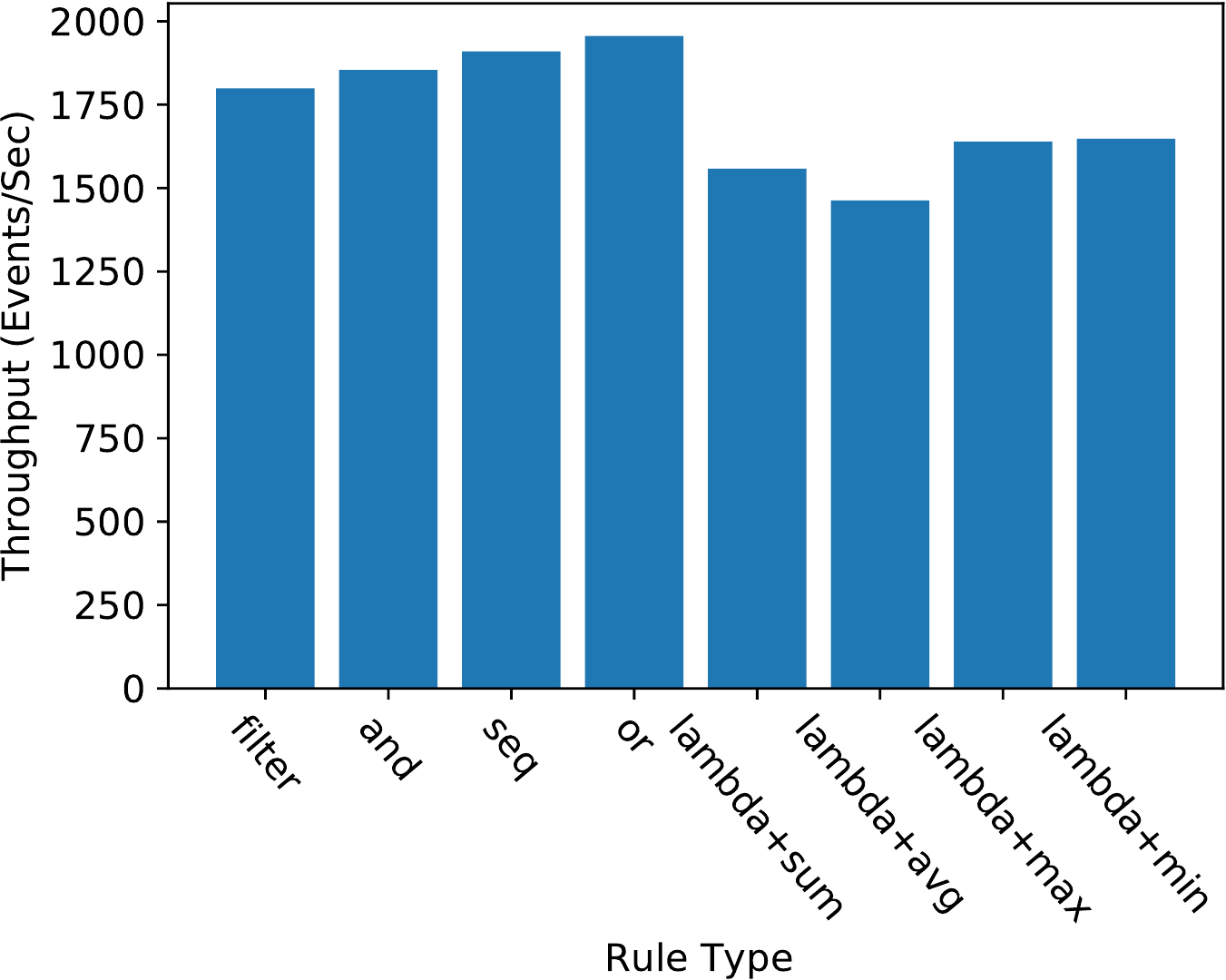}
  \caption{Throughput Test of the micro CEP Engine (Throughput vs Rule Type).}
  \label{Throughput_2}
\end{figure}

Figure \ref{Throughput_1} presents the result of the throughput test. We evaluated the average throughput for each setting by measuring how much time the engine needs for processing a stream of 10000 sensor events. By varying the number of rules in the engine, we estimated the relationship between the throughput and the number of rules. We constructed events intentionally so that each event fires each rule exactly once. After the event gets processed by all the rules, it is effectively discarded by the engine. We ran a detailed investigation of the number of rules from one to five. Usually, only a few rules are implemented on one constrained tiny IoT device due to low energy consumption and limited computational resources. We also observed that many industrial use cases (including the one from the last section) could be achieved with a load of up to five rules per sensor node. Nevertheless, we extended the test for more demanding industrial applications where the number of rules are increased up to twenty. The result shows that the more rules we add, the less throughput we obtain. The throughput is not linearly correlated to the number of rules as not each rule added requires the same computational resources. 


As shown in Figure \ref{Throughput_2}, we tested the throughput performance of several frequently used operators. We obtained a throughput of over 1450 events/s for every operator, which is a result that fulfills the requirements of most distributed IIoT applications. In the majority of our IIoT applications with a decentralized setting, the sensor node produces events with a frequency from 1 to 200 events/second. The lower frequency is desired with slower changing quantities such as temperature (\char`\~1 event/second), whereas higher frequency is needed when we measure, e.g., vibration (\char`\~200 events/second). This test shows that the micro CEP engine running on a constrained MCU could easily handle such a workload. It is worth noting that the micro CEP engine can also run on non-constrained devices (not bare metal devices), such as Siemens SIMATIC IOT2050\footnote{https://new.siemens.com/global/en/products/automation/pc-based/iot-gateways/simatic-iot2050.html} or Raspberry Pi, if higher throughputs are demanded.

\section{CONCLUSION AND FUTURE WORK}
\label{section:conclusion}

In this work, we propose a novel framework targeted for constrained field devices, which incorporates the inference ability of TinyML and the power of micro CEP to detect diversified complex events in distributed sensor networks. 

The core building blocks of the framework are the TinyOL system and the micro CEP engine. By leveraging the TinyOL system, we can run NNs on constrained tiny IIoT devices and improve existing pre-built models by learning from the field data.  The micro CEP engine is able to capture the temporal and spatial relationship between primitive events directly at the edge. In particular, the user-friendly CEP grammar supports users with no programming knowledge to configure the reasoning patterns at any time. The TinyML and the CEP's synergy play an important role in enabling versatile and flexible IIoT where they compensate one's drawback by acquiring other's strengths. Moreover, we construct an industrial use case of machine safety monitoring to illustrate the framework's functionality and evaluate its performance statistically.  By validating the concept on two Arduino boards with a cooling ventilator, we proved the feasibility, flexibility, and effectiveness of the proposed solution. At last, we evaluated the framework's performance regarding runtime, memory consumption, inference speed, and throughput.

The future work involves the following: (1) The current solution processes all the data with a deterministic model. However, the uncertainty is the nature of the NN model. Thus, we will include probability in the reasoning progress of our system. (2) We will provide a more user-friendly interface to manage program topology. Ideally, the user might "drag and drop" rules and models to configure the system. (3) We will realize the mass management of devices in IIoT networks. (4) We will improve the performance of the framework.

\bibliographystyle{ACM-Reference-Format}
\bibliography{sample-base}


\begin{thebibliography}{41}


\ifx \showCODEN    \undefined \def \showCODEN     #1{\unskip}     \fi
\ifx \showDOI      \undefined \def \showDOI       #1{#1}\fi
\ifx \showISBNx    \undefined \def \showISBNx     #1{\unskip}     \fi
\ifx \showISBNxiii \undefined \def \showISBNxiii  #1{\unskip}     \fi
\ifx \showISSN     \undefined \def \showISSN      #1{\unskip}     \fi
\ifx \showLCCN     \undefined \def \showLCCN      #1{\unskip}     \fi
\ifx \shownote     \undefined \def \shownote      #1{#1}          \fi
\ifx \showarticletitle \undefined \def \showarticletitle #1{#1}   \fi
\ifx \showURL      \undefined \def \showURL       {\relax}        \fi
\providecommand\bibfield[2]{#2}
\providecommand\bibinfo[2]{#2}
\providecommand\natexlab[1]{#1}
\providecommand\showeprint[2][]{arXiv:#2}

\bibitem[\protect\citeauthoryear{Amazon}{Amazon}{2021}]%
        {ASW21}
\bibfield{author}{\bibinfo{person}{Amazon}.} \bibinfo{year}{2021}\natexlab{}.
\newblock \bibinfo{title}{Amazon Kinesis Data Analytics}.
\newblock
\newblock
\urldef\tempurl%
\url{https://aws.amazon.com/kinesis/data-analytics}
\showURL{%
Retrieved Feb 27, 2021 from \tempurl}


\bibitem[\protect\citeauthoryear{Anicic, Rudolph, Fodor, and Stojanovic}{Anicic
  et~al\mbox{.}}{2012}]%
        {DarkoAnicic2012}
\bibfield{author}{\bibinfo{person}{Darko Anicic}, \bibinfo{person}{Sebastian
  Rudolph}, \bibinfo{person}{Paul Fodor}, {and} \bibinfo{person}{Nenad
  Stojanovic}.} \bibinfo{year}{2012}\natexlab{}.
\newblock \showarticletitle{Stream Reasoning and Complex Event Processing in
  ETALIS}.
\newblock \bibinfo{journal}{\emph{Semantic Web}} \bibinfo{volume}{3},
  \bibinfo{number}{4} (\bibinfo{year}{2012}), \bibinfo{pages}{397--407}.
\newblock
\urldef\tempurl%
\url{https://doi.org/10.3233/sw-2011-0053}
\showDOI{\tempurl}


\bibitem[\protect\citeauthoryear{Arduino}{Arduino}{2021}]%
        {Arduino21}
\bibfield{author}{\bibinfo{person}{Arduino}.} \bibinfo{year}{2021}\natexlab{}.
\newblock \bibinfo{title}{Arduino Nano 33 BLE Sense}.
\newblock
\newblock
\urldef\tempurl%
\url{https://store.arduino.cc/arduino-nano-33-ble-sense}
\showURL{%
Retrieved Feb 27, 2021 from \tempurl}


\bibitem[\protect\citeauthoryear{Bruns, Dunkel, and Offel}{Bruns
  et~al\mbox{.}}{2019}]%
        {RalfBruns2019}
\bibfield{author}{\bibinfo{person}{Ralf Bruns}, \bibinfo{person}{J{\"u}rgen
  Dunkel}, {and} \bibinfo{person}{Norman Offel}.}
  \bibinfo{year}{2019}\natexlab{}.
\newblock \showarticletitle{Learning of Complex Event Processing Rules with
  Genetic Programming}.
\newblock \bibinfo{journal}{\emph{Expert Systems with Applications}}
  \bibinfo{volume}{129} (\bibinfo{year}{2019}), \bibinfo{pages}{186--199}.
\newblock
\urldef\tempurl%
\url{https://doi.org/10.1016/j.eswa.2019.04.007}
\showDOI{\tempurl}


\bibitem[\protect\citeauthoryear{Chen, Fu, Sung, and Wang}{Chen
  et~al\mbox{.}}{2014}]%
        {ChingyuChen2014}
\bibfield{author}{\bibinfo{person}{Chingyu Chen}, \bibinfo{person}{JuiHsi Fu},
  \bibinfo{person}{Today Sung}, {and} \bibinfo{person}{Ping-Feng Wang}.}
  \bibinfo{year}{2014}\natexlab{}.
\newblock \showarticletitle{Complex Event Processing for the Internet of Things
  and Its Applications}. In \bibinfo{booktitle}{\emph{2014 IEEE International
  Conference on Automation Science and Engineering (CASE)}}.
  \bibinfo{publisher}{IEEE}, \bibinfo{address}{New Taipei, Taiwan},
  \bibinfo{pages}{1144--1149}.
\newblock
\urldef\tempurl%
\url{https://doi.org/10.1109/CoASE.2014.6899470}
\showDOI{\tempurl}


\bibitem[\protect\citeauthoryear{de~Prado~et al.}{de~Prado~et al.}{2020}]%
        {Miguel2020}
\bibfield{author}{\bibinfo{person}{Miguel de~Prado~et al.}}
  \bibinfo{year}{2020}\natexlab{}.
\newblock \bibinfo{title}{Robust Navigation with tinyML for Autonomous
  Mini-Vehicles}.
\newblock
\newblock
\showeprint[arxiv]{2007.00302}~[cs.LG]


\bibitem[\protect\citeauthoryear{{Dias de Assunção}, {da Silva Veith}, and
  Buyya}{{Dias de Assunção} et~al\mbox{.}}{2018}]%
        {MarcosDiasdeAssuncao2018}
\bibfield{author}{\bibinfo{person}{Marcos {Dias de Assunção}},
  \bibinfo{person}{Alexandre {da Silva Veith}}, {and} \bibinfo{person}{Rajkumar
  Buyya}.} \bibinfo{year}{2018}\natexlab{}.
\newblock \showarticletitle{Distributed Data Stream Processing and Edge
  Computing: A Survey on Resource Elasticity and Future Directions}.
\newblock \bibinfo{journal}{\emph{Journal of Network and Computer
  Applications}}  \bibinfo{volume}{103} (\bibinfo{year}{2018}),
  \bibinfo{pages}{1--17}.
\newblock
\showISSN{1084-8045}
\urldef\tempurl%
\url{https://doi.org/10.1016/j.jnca.2017.12.001}
\showDOI{\tempurl}


\bibitem[\protect\citeauthoryear{et~al.}{et~al.}{2021}]%
        {hogan2021}
\bibfield{author}{\bibinfo{person}{Aidan~Hogan et al.}}
  \bibinfo{year}{2021}\natexlab{}.
\newblock \bibinfo{title}{Knowledge Graphs}.
\newblock
\newblock
\showeprint[arxiv]{2003.02320}~[cs.AI]


\bibitem[\protect\citeauthoryear{et~al.}{et~al.}{2019a}]%
        {NEURIPS2019_bdbca288}
\bibfield{author}{\bibinfo{person}{Adam~Paszke et al.}}
  \bibinfo{year}{2019}\natexlab{a}.
\newblock \showarticletitle{PyTorch: An Imperative Style, High-Performance Deep
  Learning Library}. In \bibinfo{booktitle}{\emph{Advances in Neural
  Information Processing Systems}}, Vol.~\bibinfo{volume}{32}.
  \bibinfo{publisher}{Curran Associates, Inc.}, \bibinfo{address}{Canada}.
\newblock
\urldef\tempurl%
\url{https://proceedings.neurips.cc/paper/2019/file/bdbca288fee7f92f2bfa9f7012727740-Paper.pdf}
\showURL{%
\tempurl}


\bibitem[\protect\citeauthoryear{et~al.}{et~al.}{2018a}]%
        {Bert2018}
\bibfield{author}{\bibinfo{person}{Bert~Moons et al.}}
  \bibinfo{year}{2018}\natexlab{a}.
\newblock \showarticletitle{BinarEye: An Always-on Energy-Accuracy-Scalable
  Binary CNN Processor with All Memory on Chip in 28nm CMOS}. In
  \bibinfo{booktitle}{\emph{2018 IEEE Custom Integrated Circuits Conference
  (CICC)}}. IEEE, \bibinfo{publisher}{IEEE}, \bibinfo{address}{San Diego, CA,
  USA}, \bibinfo{pages}{1--4}.
\newblock
\urldef\tempurl%
\url{https://doi.org/10.1109/cicc.2018.8357071}
\showDOI{\tempurl}


\bibitem[\protect\citeauthoryear{et~al.}{et~al.}{2019b}]%
        {Jinook2019}
\bibfield{author}{\bibinfo{person}{Jinook~Song et al.}}
  \bibinfo{year}{2019}\natexlab{b}.
\newblock \showarticletitle{7.1 an 11.5 tops/w 1024-mac butterfly structure
  dual-core sparsity-aware neural processing unit in 8nm flagship mobile soc}.
  In \bibinfo{booktitle}{\emph{2019 IEEE International Solid-State Circuits
  Conference (ISSCC)}}. IEEE, \bibinfo{publisher}{IEEE}, \bibinfo{address}{San
  Francisco, CA, USA}, \bibinfo{pages}{130--132}.
\newblock
\urldef\tempurl%
\url{https://doi.org/10.1109/isscc.2019.8662476}
\showDOI{\tempurl}


\bibitem[\protect\citeauthoryear{et~al.}{et~al.}{2007}]%
        {Lars2007}
\bibfield{author}{\bibinfo{person}{Lars~Brenna et al.}}
  \bibinfo{year}{2007}\natexlab{}.
\newblock \showarticletitle{Cayuga: A High-Performance Event Processing
  Engine}. In \bibinfo{booktitle}{\emph{Proceedings of the 2007 ACM SIGMOD
  International Conference on Management of Data}} (Beijing, China)
  \emph{(\bibinfo{series}{SIGMOD '07})}. \bibinfo{publisher}{Association for
  Computing Machinery}, \bibinfo{address}{New York, NY, USA},
  \bibinfo{pages}{1100--1102}.
\newblock
\showISBNx{9781595936868}
\urldef\tempurl%
\url{https://doi.org/10.1145/1247480.1247620}
\showDOI{\tempurl}


\bibitem[\protect\citeauthoryear{et~al.}{et~al.}{2010}]%
        {Lajos2010}
\bibfield{author}{\bibinfo{person}{Lajos Jen{\H{o}}~F{\"u}l{\"o}p et al.}}
  \bibinfo{year}{2010}\natexlab{}.
\newblock \showarticletitle{Survey on Complex Event Processing and Predictive
  Analytics}. In \bibinfo{booktitle}{\emph{Proceedings of the Fifth Balkan
  Conference in Informatics}}. Citeseer, \bibinfo{publisher}{Association for
  Computing Machinery}, \bibinfo{address}{New York, NY, USA},
  \bibinfo{pages}{26--31}.
\newblock


\bibitem[\protect\citeauthoryear{et~al.}{et~al.}{2019c}]%
        {Lina2019}
\bibfield{author}{\bibinfo{person}{Lina~Lan et al.}}
  \bibinfo{year}{2019}\natexlab{c}.
\newblock \showarticletitle{A Universal Complex Event Processing Mechanism
  Based on Edge Computing for Internet of Things Real-Time Monitoring}.
\newblock \bibinfo{journal}{\emph{IEEE Access}}  \bibinfo{volume}{7}
  (\bibinfo{year}{2019}), \bibinfo{pages}{101865--101878}.
\newblock
\urldef\tempurl%
\url{https://doi.org/10.1109/ACCESS.2019.2930313}
\showDOI{\tempurl}


\bibitem[\protect\citeauthoryear{et~al.}{et~al.}{2015a}]%
        {tensorflow2015-whitepaper}
\bibfield{author}{\bibinfo{person}{Mart\'{\i}n~Abadi et al.}}
  \bibinfo{year}{2015}\natexlab{a}.
\newblock \bibinfo{title}{{TensorFlow}: Large-Scale Machine Learning on
  Heterogeneous Systems}.
\newblock
\newblock
\urldef\tempurl%
\url{https://www.tensorflow.org/}
\showURL{%
\tempurl}
\newblock
\shownote{Software available from tensorflow.org.}


\bibitem[\protect\citeauthoryear{et~al.}{et~al.}{2015b}]%
        {Nijat2015}
\bibfield{author}{\bibinfo{person}{Nijat~Mehdiyev et al.}}
  \bibinfo{year}{2015}\natexlab{b}.
\newblock \showarticletitle{Determination of Rule Patterns in Complex Event
  Processing Using Machine Learning Techniques}.
\newblock \bibinfo{journal}{\emph{Procedia Computer Science}}
  \bibinfo{volume}{61} (\bibinfo{year}{2015}), \bibinfo{pages}{395--401}.
\newblock
\urldef\tempurl%
\url{https://doi.org/10.1016/j.procs.2015.09.168}
\showDOI{\tempurl}


\bibitem[\protect\citeauthoryear{et~al.}{et~al.}{2018b}]%
        {Pablo2018}
\bibfield{author}{\bibinfo{person}{Pablo~Graubner et al.}}
  \bibinfo{year}{2018}\natexlab{b}.
\newblock \showarticletitle{Multimodal Complex Event Processing on Mobile
  Devices}. In \bibinfo{booktitle}{\emph{Proceedings of the 12th ACM
  International Conference on Distributed and Event-Based Systems}} (Hamilton,
  New Zealand) \emph{(\bibinfo{series}{DEBS '18})}.
  \bibinfo{publisher}{Association for Computing Machinery},
  \bibinfo{address}{New York, NY, USA}, \bibinfo{pages}{112--123}.
\newblock
\showISBNx{9781450357821}
\urldef\tempurl%
\url{https://doi.org/10.1145/3210284.3210289}
\showDOI{\tempurl}


\bibitem[\protect\citeauthoryear{et~al.}{et~al.}{2020}]%
        {Robert2020}
\bibfield{author}{\bibinfo{person}{Robert~David et al.}}
  \bibinfo{year}{2020}\natexlab{}.
\newblock \bibinfo{title}{Tensorflow Lite Micro: Embedded Machine Learning on
  tinyML Systems}.
\newblock
\newblock
\showeprint[arxiv]{1801.06601}~[cs.LG]


\bibitem[\protect\citeauthoryear{et~al.}{et~al.}{2018c}]%
        {Tianqi2018}
\bibfield{author}{\bibinfo{person}{Tianqi~Chen et al.}}
  \bibinfo{year}{2018}\natexlab{c}.
\newblock \showarticletitle{TVM: An Automated End-to-End Optimizing Compiler
  for Deep Learning}. In \bibinfo{booktitle}{\emph{13th USENIX Symposium on
  Operating Systems Design and Implementation (OSDI 18)}}.
  \bibinfo{publisher}{USENIX Association}, \bibinfo{address}{USA},
  \bibinfo{pages}{578--594}.
\newblock


\bibitem[\protect\citeauthoryear{et~al.}{et~al.}{2019d}]%
        {Tianwei2019}
\bibfield{author}{\bibinfo{person}{Tianwei~Xing et al.}}
  \bibinfo{year}{2019}\natexlab{d}.
\newblock \showarticletitle{Deepcep: Deep Complex Event Processing Using
  Distributed Multimodal Information}. In \bibinfo{booktitle}{\emph{2019 IEEE
  International Conference on Smart Computing (SMARTCOMP)}}.
  \bibinfo{publisher}{IEEE}, \bibinfo{address}{Washington, DC, USA},
  \bibinfo{pages}{87--92}.
\newblock
\urldef\tempurl%
\url{https://doi.org/10.1109/smartcomp.2019.00034}
\showDOI{\tempurl}


\bibitem[\protect\citeauthoryear{Foundation}{Foundation}{2021}]%
        {tinyML2021}
\bibfield{author}{\bibinfo{person}{TinyML Foundation}.}
  \bibinfo{year}{2021}\natexlab{}.
\newblock \bibinfo{title}{tinyML.org}.
\newblock
\newblock
\urldef\tempurl%
\url{https://www.tinyml.org}
\showURL{%
Retrieved Feb 27, 2021 from \tempurl}


\bibitem[\protect\citeauthoryear{Fundation}{Fundation}{2021}]%
        {apache21}
\bibfield{author}{\bibinfo{person}{The Apache~Software Fundation}.}
  \bibinfo{year}{2021}\natexlab{}.
\newblock \bibinfo{title}{Apache Flink}.
\newblock
\newblock
\urldef\tempurl%
\url{https://ci.apache.org/projects/flink/flink-docs-release-1.12}
\showURL{%
Retrieved Feb 27, 2021 from \tempurl}


\bibitem[\protect\citeauthoryear{Giatrakos, Alevizos, Artikis, Deligiannakis,
  and Garofalakis}{Giatrakos et~al\mbox{.}}{2020}]%
        {Giatrakos2020}
\bibfield{author}{\bibinfo{person}{Nikos Giatrakos}, \bibinfo{person}{Elias
  Alevizos}, \bibinfo{person}{Alexander Artikis}, \bibinfo{person}{Antonios
  Deligiannakis}, {and} \bibinfo{person}{Minos Garofalakis}.}
  \bibinfo{year}{2020}\natexlab{}.
\newblock \showarticletitle{Complex Event Recognition in the Big Data Era: A
  Survey}.
\newblock \bibinfo{journal}{\emph{The VLDB Journal}}  \bibinfo{volume}{29}
  (\bibinfo{date}{01} \bibinfo{year}{2020}).
\newblock
\urldef\tempurl%
\url{https://doi.org/10.1007/s00778-019-00557-w}
\showDOI{\tempurl}


\bibitem[\protect\citeauthoryear{Google}{Google}{2021}]%
        {TFLiteMicro21}
\bibfield{author}{\bibinfo{person}{Google}.} \bibinfo{year}{2021}\natexlab{}.
\newblock \bibinfo{title}{Tensorflow Lite Micro Guide}.
\newblock
\newblock
\urldef\tempurl%
\url{https://www.tensorflow.org/lite/microcontrollers}
\showURL{%
Retrieved Feb 27, 2021 from \tempurl}


\bibitem[\protect\citeauthoryear{Govindarajan, Simmhan, Jamadagni, and
  Misra}{Govindarajan et~al\mbox{.}}{2014}]%
        {NithyashriGovindarajan2014}
\bibfield{author}{\bibinfo{person}{Nithyashri Govindarajan},
  \bibinfo{person}{Yogesh Simmhan}, \bibinfo{person}{Nitin Jamadagni}, {and}
  \bibinfo{person}{Prasant Misra}.} \bibinfo{year}{2014}\natexlab{}.
\newblock \showarticletitle{Event Processing Across Edge and the Cloud for
  Internet of Things Applications}. In \bibinfo{booktitle}{\emph{Proceedings of
  the 20th International Conference on Management of Data}} (Hyderabad, India)
  \emph{(\bibinfo{series}{COMAD '14})}. \bibinfo{publisher}{Computer Society of
  India}, \bibinfo{address}{Mumbai, Maharashtra, IND},
  \bibinfo{pages}{101--104}.
\newblock


\bibitem[\protect\citeauthoryear{Han, Mao, and Dally}{Han
  et~al\mbox{.}}{2015}]%
        {SonHan2015}
\bibfield{author}{\bibinfo{person}{Son Han}, \bibinfo{person}{Huizi Mao}, {and}
  \bibinfo{person}{William~J. Dally}.} \bibinfo{year}{2015}\natexlab{}.
\newblock \bibinfo{title}{Deep Compression: Compressing Deep Neural Networks
  with Pruning, Trained Quantization and Huffman Coding}.
\newblock
\newblock
\showeprint[arxiv]{1510.00149}~[cs.LG]


\bibitem[\protect\citeauthoryear{Lai, Suda, and Chandra}{Lai
  et~al\mbox{.}}{2018}]%
        {LiangzhenLai2018}
\bibfield{author}{\bibinfo{person}{Liangzhen Lai}, \bibinfo{person}{Naveen
  Suda}, {and} \bibinfo{person}{Vikas Chandra}.}
  \bibinfo{year}{2018}\natexlab{}.
\newblock \bibinfo{title}{Cmsis-nn: Efficient Neural Network Kernels for Arm
  Cortex-M CPUs}.
\newblock
\newblock
\showeprint[arxiv]{1801.06601}~[cs.LG]


\bibitem[\protect\citeauthoryear{Mohamed and Al-Jaroodi}{Mohamed and
  Al-Jaroodi}{2014}]%
        {Mohamed2014}
\bibfield{author}{\bibinfo{person}{Nader Mohamed} {and}
  \bibinfo{person}{Jameela Al-Jaroodi}.} \bibinfo{year}{2014}\natexlab{}.
\newblock \showarticletitle{Real-time big data analytics: Applications and
  challenges}. In \bibinfo{booktitle}{\emph{2014 International Conference on
  High Performance Computing Simulation (HPCS)}}. \bibinfo{publisher}{IEEE},
  \bibinfo{address}{Bologna, Italy}, \bibinfo{pages}{305--310}.
\newblock
\urldef\tempurl%
\url{https://doi.org/10.1109/HPCSim.2014.6903700}
\showDOI{\tempurl}


\bibitem[\protect\citeauthoryear{Mousheimish, Taher, and Zeitouni}{Mousheimish
  et~al\mbox{.}}{2017}]%
        {RaefMousheimish2017}
\bibfield{author}{\bibinfo{person}{Raef Mousheimish}, \bibinfo{person}{Yehia
  Taher}, {and} \bibinfo{person}{Karine Zeitouni}.}
  \bibinfo{year}{2017}\natexlab{}.
\newblock \showarticletitle{Automatic Learning of Predictive CEP Rules:
  Bridging the Gap Between Data Mining and Complex Event Processing}. In
  \bibinfo{booktitle}{\emph{Proceedings of the 11th ACM International
  Conference on Distributed and Event-based Systems}}.
  \bibinfo{publisher}{Association for Computing Machinery},
  \bibinfo{address}{New York, NY, USA}, \bibinfo{pages}{158--169}.
\newblock
\urldef\tempurl%
\url{https://doi.org/10.1145/3093742.3093917}
\showDOI{\tempurl}


\bibitem[\protect\citeauthoryear{OmniVision}{OmniVision}{2021}]%
        {camera2021}
\bibfield{author}{\bibinfo{person}{OmniVision}.}
  \bibinfo{year}{2021}\natexlab{}.
\newblock \bibinfo{title}{OV7675 Image Sensor}.
\newblock
\newblock
\urldef\tempurl%
\url{https://www.ovt.com/sensors/OV7675}
\showURL{%
Retrieved Feb 27, 2021 from \tempurl}


\bibitem[\protect\citeauthoryear{Power and Kotonya}{Power and Kotonya}{2019}]%
        {Power2019}
\bibfield{author}{\bibinfo{person}{Alexander Power} {and}
  \bibinfo{person}{Gerald Kotonya}.} \bibinfo{year}{2019}\natexlab{}.
\newblock \showarticletitle{Providing Fault Tolerance via Complex Event
  Processing and Machine Learning for IoT Systems}. In
  \bibinfo{booktitle}{\emph{Proceedings of the 9th International Conference on
  the Internet of Things}}. \bibinfo{publisher}{Association for Computing
  Machinery}, \bibinfo{address}{New York, NY, USA}, \bibinfo{pages}{1--7}.
\newblock
\urldef\tempurl%
\url{https://doi.org/10.1145/3365871.3365872}
\showDOI{\tempurl}


\bibitem[\protect\citeauthoryear{Proctor}{Proctor}{2012}]%
        {Proctor2012}
\bibfield{author}{\bibinfo{person}{Mark Proctor}.}
  \bibinfo{year}{2012}\natexlab{}.
\newblock \showarticletitle{Drools: A Rule Engine for Complex Event
  Processing}. In \bibinfo{booktitle}{\emph{Applications of Graph
  Transformations with Industrial Relevance}},
  \bibfield{editor}{\bibinfo{person}{Andy Sch{\"u}rr} {and}
  \bibinfo{person}{D{\'a}niel Varr{\'o}}} (Eds.). \bibinfo{publisher}{Springer
  Berlin Heidelberg}, \bibinfo{address}{Berlin, Heidelberg},
  \bibinfo{pages}{2--2}.
\newblock
\showISBNx{978-3-642-34176-2}
\urldef\tempurl%
\url{https://doi.org/10.1007/978-3-642-34176-2_2}
\showDOI{\tempurl}


\bibitem[\protect\citeauthoryear{Rastegari, Ordonez, Redmon, and
  Farhadi}{Rastegari et~al\mbox{.}}{2016}]%
        {MohammadRastegari2016}
\bibfield{author}{\bibinfo{person}{Mohammad Rastegari},
  \bibinfo{person}{Vicente Ordonez}, \bibinfo{person}{Joseph Redmon}, {and}
  \bibinfo{person}{Ali Farhadi}.} \bibinfo{year}{2016}\natexlab{}.
\newblock \showarticletitle{Xnor-Net: Imagenet Classification Using Binary
  Convolutional Neural Networks}. In \bibinfo{booktitle}{\emph{European
  Conference on Computer Vision}}. Springer, \bibinfo{publisher}{Springer,
  Cham}, \bibinfo{address}{Amsterdam, The Netherlands},
  \bibinfo{pages}{525--542}.
\newblock
\urldef\tempurl%
\url{https://doi.org/10.1007/978-3-319-46493-0_32}
\showDOI{\tempurl}


\bibitem[\protect\citeauthoryear{Ren, Anicic, and Runkler}{Ren
  et~al\mbox{.}}{2021}]%
        {Ren2021}
\bibfield{author}{\bibinfo{person}{Haoyu Ren}, \bibinfo{person}{Darko Anicic},
  {and} \bibinfo{person}{Thomas Runkler}.} \bibinfo{year}{2021}\natexlab{}.
\newblock \bibinfo{title}{TinyOL: TinyML with Online-Learning on
  Microcontrollers}.
\newblock
\newblock
\showeprint[arxiv]{2103.08295}~[cs.LG]


\bibitem[\protect\citeauthoryear{Sanchez-Iborra and Skarmeta}{Sanchez-Iborra
  and Skarmeta}{2020}]%
        {SanchezIborra2020}
\bibfield{author}{\bibinfo{person}{Ramon Sanchez-Iborra} {and}
  \bibinfo{person}{Antonio~F. Skarmeta}.} \bibinfo{year}{2020}\natexlab{}.
\newblock \showarticletitle{TinyML-Enabled Frugal Smart Objects: Challenges and
  Opportunities}.
\newblock \bibinfo{journal}{\emph{IEEE Circuits and Systems Magazine}}
  \bibinfo{volume}{20}, \bibinfo{number}{3} (\bibinfo{year}{2020}),
  \bibinfo{pages}{4--18}.
\newblock
\urldef\tempurl%
\url{https://doi.org/10.1109/mcas.2020.3005467}
\showDOI{\tempurl}


\bibitem[\protect\citeauthoryear{Soto, Jentsch, Preuveneers, and
  Ilie-Zudor}{Soto et~al\mbox{.}}{2016}]%
        {JoseAngelCarvajalSoto2016}
\bibfield{author}{\bibinfo{person}{Jos{\'e} Angel~Carvajal Soto},
  \bibinfo{person}{Marc Jentsch}, \bibinfo{person}{Davy Preuveneers}, {and}
  \bibinfo{person}{Elisabeth Ilie-Zudor}.} \bibinfo{year}{2016}\natexlab{}.
\newblock \showarticletitle{CEML: Mixing and Moving Complex Event Processing
  and Machine Learning to the Edge of the Network for IoT Applications}. In
  \bibinfo{booktitle}{\emph{Proceedings of the 6th International Conference on
  the Internet of Things}}. \bibinfo{publisher}{Association for Computing
  Machinery}, \bibinfo{address}{New York, NY, USA}, \bibinfo{pages}{103--110}.
\newblock
\urldef\tempurl%
\url{https://doi.org/10.1145/2991561.2991575}
\showDOI{\tempurl}


\bibitem[\protect\citeauthoryear{STMicroelectronics}{STMicroelectronics}{2021}]%
        {STM21}
\bibfield{author}{\bibinfo{person}{STMicroelectronics}.}
  \bibinfo{year}{2021}\natexlab{}.
\newblock \bibinfo{title}{AI expansion pack for STM32CubeMX}.
\newblock
\newblock
\urldef\tempurl%
\url{https://www.st.com/en/embedded-software/x-cube-ai.html}
\showURL{%
Retrieved Feb 27, 2021 from \tempurl}


\bibitem[\protect\citeauthoryear{TIBCO}{TIBCO}{2021}]%
        {TIBCO21}
\bibfield{author}{\bibinfo{person}{TIBCO}.} \bibinfo{year}{2021}\natexlab{}.
\newblock \bibinfo{title}{TIBCO}.
\newblock
\newblock
\urldef\tempurl%
\url{https://www.tibco.com}
\showURL{%
Retrieved Feb 27, 2021 from \tempurl}


\bibitem[\protect\citeauthoryear{Vrbaski, Bolic, and Majumdar}{Vrbaski
  et~al\mbox{.}}{2018}]%
        {MiraVrbaski2018}
\bibfield{author}{\bibinfo{person}{Mira Vrbaski}, \bibinfo{person}{Miodrag
  Bolic}, {and} \bibinfo{person}{Shikharesh Majumdar}.}
  \bibinfo{year}{2018}\natexlab{}.
\newblock \showarticletitle{Complex Event Recognition Notification Methodology
  for Uncertain IoT Systems Based on Micro-Service Architecture}. In
  \bibinfo{booktitle}{\emph{2018 IEEE 6th International Conference on Future
  Internet of Things and Cloud (FiCloud)}}. \bibinfo{publisher}{IEEE},
  \bibinfo{address}{Barcelona, Spain}, \bibinfo{pages}{184--191}.
\newblock
\urldef\tempurl%
\url{https://doi.org/10.1109/FiCloud.2018.00034}
\showDOI{\tempurl}


\bibitem[\protect\citeauthoryear{Wang, Gao, and Chen}{Wang
  et~al\mbox{.}}{2018}]%
        {YonghengWang2018}
\bibfield{author}{\bibinfo{person}{Yongheng Wang}, \bibinfo{person}{Hui Gao},
  {and} \bibinfo{person}{Guidan Chen}.} \bibinfo{year}{2018}\natexlab{}.
\newblock \showarticletitle{Predictive Complex Event Processing Based on
  Evolving Bayesian Networks}.
\newblock \bibinfo{journal}{\emph{Pattern Recognition Letters}}
  \bibinfo{volume}{105} (\bibinfo{year}{2018}), \bibinfo{pages}{207--216}.
\newblock
\urldef\tempurl%
\url{https://doi.org/10.1016/j.patrec.2017.05.008}
\showDOI{\tempurl}


\bibitem[\protect\citeauthoryear{Wanner, Wissuchek, and Janiesch}{Wanner
  et~al\mbox{.}}{2019}]%
        {JonasWanner2019}
\bibfield{author}{\bibinfo{person}{Jonas Wanner}, \bibinfo{person}{Christopher
  Wissuchek}, {and} \bibinfo{person}{Christian Janiesch}.}
  \bibinfo{year}{2019}\natexlab{}.
\newblock \showarticletitle{Machine Learning and Complex Event Processing. A
  Review of Real-time Data Analytics for the Industrial Internet of Things}.
\newblock \bibinfo{journal}{\emph{Enterprise Modelling and Information Systems
  Architectures (EMISAJ) – International Journal of Conceptual Modeling}}
  \bibinfo{volume}{15}, \bibinfo{number}{1} (\bibinfo{year}{2019}),
  \bibinfo{pages}{1--27}.
\newblock
\urldef\tempurl%
\url{https://doi.org/10.18417/emisa.15.1}
\showDOI{\tempurl}


\end{thebibliography}


\end{document}